\documentclass[sigconf]{acmart}

\renewcommand\footnotetextcopyrightpermission[1]{}
\usepackage{algorithm}  
\usepackage{algpseudocode}  
\usepackage{amsmath}  
\usepackage{multirow}
\usepackage{multicol}
\usepackage{pifont} 
\usepackage{balance}

\AtBeginDocument{%
  }

\copyrightyear{2025} 
\acmYear{2025} 
\setcopyright{cc}
\setcctype{by-nc-sa}
\setcctype{by-nc-sa}
\acmConference[CIKM '25]{Proceedings of the 34th ACM International
Conference on Information and Knowledge Management}{November 10--14,
2025}{Seoul, Republic of Korea}
\acmBooktitle{Proceedings of the 34th ACM International Conference on
Information and Knowledge Management (CIKM '25), November 10--14, 2025,
Seoul, Republic of Korea}\acmDOI{10.1145/3746252.3761146}
\acmISBN{979-8-4007-2040-6/2025/11}

\begin{document}

\title{Structure-prior Informed Diffusion Model for Graph Source Localization with Limited Data}


\author{Hongyi Chen}
\additionalaffiliation{%
  \institution{Pengcheng Laboratory}
  \city{Shenzhen}
  \country{China}
}
\affiliation{%
\department{Shenzhen Key Laboratory of Ubiquitous Data Enabling, Shenzhen International Graduate School}
  \institution{Tsinghua University}
  \city{Shenzhen}
  \country{China}
}
\email{chenhy23@mails.tsinghua.edu.cn}

\author{Jingtao Ding}
\affiliation{%
    \department{Department of Electronic Engineering}
  \institution{Tsinghua University}
  \city{Beijing}
  \country{China}
}
\email{dingjt15@tsinghua.org.cn}

\author{Xiaojun Liang}
\affiliation{%
  \institution{Pengcheng Laboratory}
  \city{Shenzhen}
  \country{China}
}
\email{liangxj@pcl.ac.cn}

\author{Yong Li}
\affiliation{%
\department{Department of Electronic Engineering}
  \institution{Tsinghua University}
  \city{Beijing}
  \country{China}
}
\email{liyong07@tsinghua.edu.cn}

\author{Xiao-Ping Zhang}
\authornote{Corresponding author}
\affiliation{%
\department{Shenzhen Key Laboratory of Ubiquitous Data Enabling, Shenzhen International Graduate School}
  \institution{Tsinghua University}
  \city{Shenzhen}
  \country{China}
}
\email{xpzhang@ieee.org}

\renewcommand{\shortauthors}{Hongyi Chen, Jingtao Ding, Xiaojun Liang, Yong Li and Xiao-Ping Zhang}

\begin{abstract}
Source localization in graph information propagation is essential for mitigating network disruptions, including misinformation spread, cyber threats, and infrastructure failures. Existing deep generative approaches face significant challenges in real-world applications due to limited propagation data availability. We present SIDSL (\textbf{S}tructure-prior \textbf{I}nformed \textbf{D}iffusion model for \textbf{S}ource \textbf{L}ocalization), a generative diffusion framework that leverages topology-aware priors to enable robust source localization with limited data. SIDSL addresses three key challenges: unknown propagation patterns through structure-based source estimations via graph label propagation, complex topology-propagation relationships via a propagation-enhanced conditional denoiser with GNN-parameterized label propagation module, and class imbalance through structure-prior biased diffusion initialization. By learning pattern-invariant features from synthetic data generated by established propagation models, SIDSL enables effective knowledge transfer to real-world scenarios. Experimental evaluation on four real-world datasets demonstrates superior performance with 7.5-13.3\% F1 score improvements over baselines, including over 19\% improvement in few-shot and 40\% in zero-shot settings, validating the framework's effectiveness for practical source localization. 
Our code can be found \href{https://github.com/tsinghua-fib-lab/SIDSL}{here}.
\end{abstract}




\maketitle

\section{Introduction}
Graph information propagation issues, such as misinformation spread, cyber threats, and infrastructure failures, have far-reaching societal consequences. Quickly identifying disruption sources is critical for impact mitigation. By analyzing snapshots of affected networks, we can trace spread origins, essential for managing disease outbreaks~\citep{ru2023inferring,ding2024comprehensive}, network security~\citep{kephart1993measuring}, and infrastructure failures~\citep{amin2007preventing}.

Early methods~\citep{lappas2010finding,shah2012rumor,prakash2012spotting,luo2013identifying,zhu2014information,zhu2014robust} for source localization in graphs rely on topology-derived metrics applicable only to specific propagation patterns like the Susceptible-Infected (SI) or Independent Cascade (IC) models. Wang et al.~\citep{wang2017multiple} introduced label propagation based on source centrality, but with limited expressiveness and scalability in encoding topological information.
Data-driven methods~\citep{dong2019multiple,wang2022invertible,hou2023sequential} overcome these limitations as they directly learn graph neural networks~(GNNs) to capture propagation processes but neglect propagation indeterminacy~\cite{ling2022source}. Recent deep generative models, including variational autoencoders~\citep{ling2022source}, normalization flows~\citep{xu2024pgsl}, and diffusion models~\citep{huang2023two,yan2024diffusion}, quantify this indeterminacy by learning empirical data distributions, advancing state-of-the-art performance.

However, collecting real-world propagation data for deep generative methods is difficult and costly, requiring models that adapt to limited data environments. This poses three main challenges.
\textbf{First, real-world graphs exhibit unknown propagation patterns that are challenging to characterize with limited data. }Existing methods~\citep{dong2019multiple,wang2022invertible,ling2022source,yan2024diffusion} rely purely on data for pattern understanding, limiting generalization to unseen scenarios.
\textbf{Second, complex interrelations between propagation patterns and graph topology are difficult to capture with limited data.} Existing methods require extensive labeled data from target networks to account for the structural heterogeneity's impact on propagation patterns. This dependence causes underperformance with limited training samples, hindering practical applicability.
\textbf{Third, inherent class imbalance between source and non-source nodes becomes more harmful under data scarcity, compromising identification accuracy.} This challenge remains unaddressed by existing solutions and requires comprehensive handling of both abundant and limited data scenarios through novel approaches that prevent model degeneration towards non-source predictions.

\begin{figure}[t]
\centering
  \includegraphics[width=\linewidth]{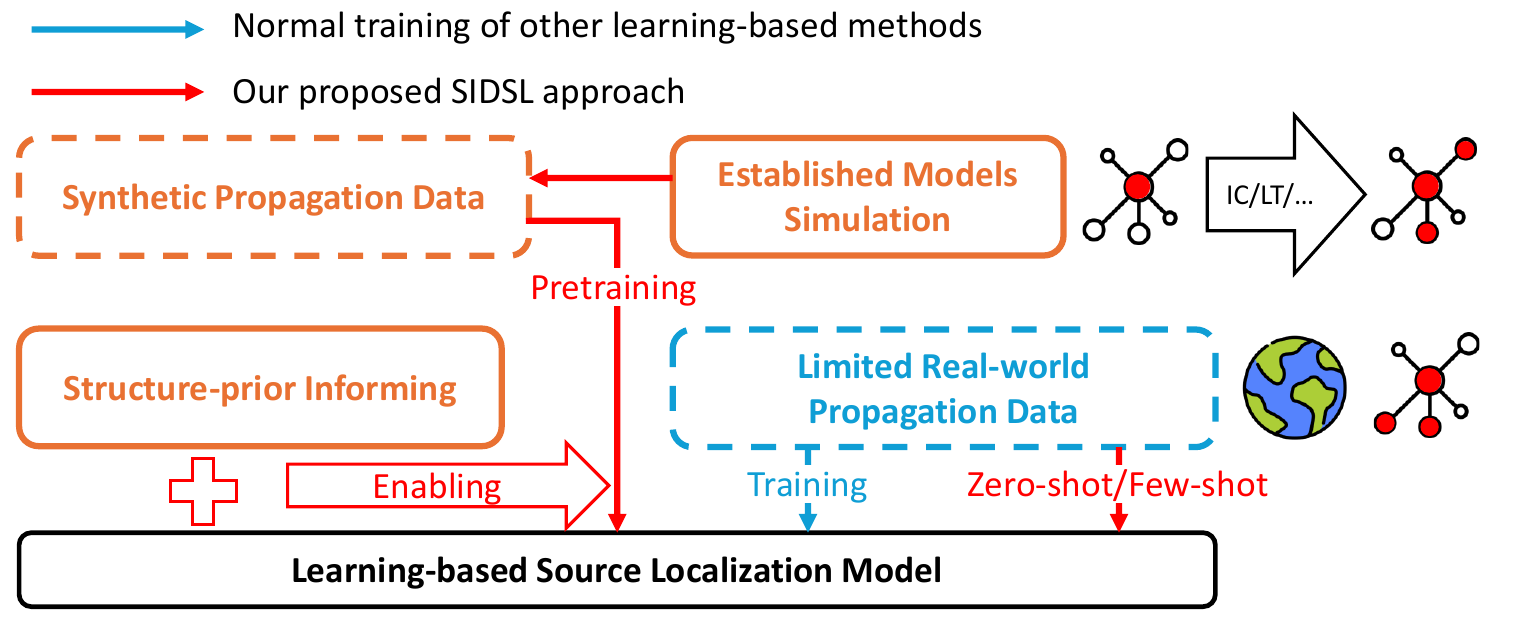} 
  \vspace{-2em}
  \caption{The proposed SIDSL approach.}
    \label{fig:approach}
    \vspace{-2em}
\end{figure}

Therefore, in this paper, we propose a novel generative diffusion framework, namely \textbf{S}tructure-prior \textbf{I}nformed \textbf{D}iffusion Model for \textbf{S}ource \textbf{L}ocalization (\textbf{SIDSL}), which leverages topology-aware priors to achieve robust source localization with limited propagation data collected from real environments. SIDSL employs a denoising diffusion model guided by structural priors to predict source node distributions from graph topology and observed node states.

To tackle the first challenge, we leverage graph structure-based source estimations, generated through graph label propagation, which identify potential sources based on locally highest infection values. Integrated into the denoising network, these topology-aware priors provide stable guidance across different propagation patterns, enhancing generalization to unseen propagation scenarios.
To address the second challenge, we design a propagation-enhanced conditional denoiser with a GNN-parameterized label propagation module (GNN-LP) that combines label propagation for tracing infection pathways with GNN for efficient topology-aware feature extraction, enabling effective learning of topology-propagation relationships from limited data.
To tackle the third challenge, we develop a structure-prior biased diffusion process that comprehensively handles the fundamental source/non-source imbalance by initializing from structure-based estimations, providing a complete solution that advances imbalance resolution in both data-rich and data-scarce scenarios.

The above designs work together in our structure-prior informed diffusion framework to capture stable structural patterns independent of specific propagation dynamics, allowing our model to learn pattern-invariant features using synthetic propagation data simulated from established models~\cite{keeling2005networks,kempe2003maximizing,kermack1927contribution,pan2025cross}. This allows effective knowledge transfer to real-world scenarios through few-shot or zero-shot learning (Figure \ref{fig:approach}).

Our contributions are summarized as follows:\\
(1)~We propose SIDSL, a structure-prior informed diffusion framework that uniquely integrates structural priors with diffusion models to effectively address the challenge of limited labeled data in source localization. Unlike existing approaches requiring abundant training data, our framework leverages domain knowledge and topology-aware priors for robust generalization to unknown propagation patterns in real-world low-data scenarios.\\
(2)~We introduce a series of innovative techniques, including a propagation-enhanced conditional denoiser with GNN-LP module and structure-prior biased denoising, which work synergistically to handle structural heterogeneity and class imbalance issues. \\
(3) We evaluate SIDSL's performance across four real-world datasets and demonstrate its superior effectiveness in source identification tasks. Our method achieves superior performance across four real-world datasets, outperforming baselines by 7.5-13.3\% in F1 scores, with over 19\% improvement in few-shot and 40\% in zero-shot settings.
Additional evaluations further validate SIDSL's effective synthetic-to-real transfer capability in improving sample efficiency and reducing training time.

\section{Related Works}
\label{r_w}
\subsection{Source Localization}
Source localization, the inverse problem of information propagation in networks, infers initial sources from diffused observations, with applications such as rumor source identification and power grid failure detection~\citep{shelke2019source,su2024rumor}. Early approaches are rule-based and rely on metrics or heuristics derived from the network’s
topology for source identification~\citep{shah2011rumors,zhu2014information,zhu2014robust}. 
For instance, \citet{shah2011rumors} developed a rumor-centrality-based maximum likelihood estimator for the Susceptible-Infected (SI)~\citep{kermack1927contribution} model. These methods, however, inadequately encode topology. Subsequently, deep learning methods emerged~\citep{lappas2010finding,luo2013identifying,wang2017multiple,dong2019multiple,wang2022invertible,hou2024new}, but most struggled to model source location uncertainty in stochastic propagation. To overcome this, deep generative models have been adopted~\cite{ling2022source,yan2024diffusion,wang2024joint,xu2024pgsl,huang2023two}.
SLVAE~\citep{ling2022source} utilizes Variational Auto-Encoders~(VAEs) and optimizes the posterior for better prediction. However, VAEs struggle with complex propagation patterns. DDMSL~\citep{yan2024diffusion} models the Susceptible-Infected-Recovered~(SIR) ~\citep{kermack1927contribution}infection process into Diffusion Models~(DM)~\citep{ho2020denoising,xu2025diffusive}, and designs a reversible residual block based on Graph Convolutional Networks~(GCNs)~\citep{kipf2016semi}. However, it requires additional intermediate propagation data and cannot be generalized to other propagation patterns. In contrast to existing methods that demand extensive training data, our novel framework leverages domain knowledge for robust localization with minimal labeled data—a significant advance for real-world scenarios. A comparison of typical methods is in Appendix~\ref{app:comp}.

\subsection{Typical Propagation Models}


Information propagation models estimate information spread in networks, applied in event prediction~\citep{zhao2021event}, adverse event detection~\citep{wang2018multi}, disease spread prediction~\citep{Tang2023Enhancing}, and so on. Two main categories exist: infection models and influence models.
Infection models such as Susceptible-Infected (SI) and Susceptible-Infected-Susceptible (SIS) manage state transitions in networks~\citep{kermack1927contribution,keeling2005networks}. Infected nodes attempt to infect neighbors with probability $\beta$, while in SIS models, infected nodes revert to susceptible with probability $\lambda$. The Susceptible-Infected-Recovered (SIR) model adds a recovered state.
Influence models include Independent Cascade (IC) and Linear Threshold (LT)~\citep{kempe2003maximizing}. In IC, newly activated nodes attempt once to activate neighbors based on edge weights. LT models activate nodes when accumulated neighbor influence exceeds a threshold.

\section{Preliminaries}
\label{Pre}
\subsection{Problem Formulation}
Our research problem is formulated as follows. Given an undirected social network $\mathcal{G} = (V, E)$ where $V$ is the node set, $E$ is the edge set, and $Y = \{Y_1, \ldots, Y_{|V|}\}$ is an infection state of all nodes in $\mathcal{G}$, which describes that a subset of nodes in $\mathcal{G}$ have been infected. Each $Y_i \in \{1, 0\}$ denotes the infection state of node $v_i \in V$, where $Y_i = 1$ indicates that $v_i$ is infected and otherwise $Y_i = 0$ indicates it is uninfected.
We aim to find the original propagation source $\hat{X}$ from the propagated observation $Y$, so that the loss with the ground Truth source set $X^* \in \{1, 0\}^{|V|\times1}$ is minimized, i.e. $\hat{X}=argmin_X||X-X^*||_2^2$. To account for the uncertainty in source localization, we need to construct a probabilistic model $P(X|Y,\mathcal{G})$, which can be used to sample for the final prediction.

\subsection{Label Propagation based Source Identification}
\label{subsec:lpsi}
In realistic situations, the intractable propagation process does not have an explicit prior, and it is also challenging to value appropriate parameters for the pre-selected underlying propagation model. To address this, \citet{wang2017multiple} propose Label Propagation based Source Identification~(LPSI). Since LPSI investigates the same problem as our research, we use it as a baseline to compare performance. LPSI captures source centrality characteristics in the method design. The centrality of sources shows that nodes far from the source are less likely to be infected than those near it~\citep{shah2012rumor}, which can also be observed in the real-world data by our analysis in the Appendix~\ref{app:assumption}. Based on these ideas, they propose to perform label propagation on the observation state of the network. By setting $Y[Y=0]=-1$ and $~\mathcal{Z}^{t=0}\xleftarrow{}Y$, the iteration of label propagation and the convergence states are as follows:
\begin{equation}
    \mathcal{Z}^{t+1}_i=\alpha\sum_{j:j\in\mathcal{N}(I)}S_{ij}\mathcal{Z}_j^t+(1-\alpha)Y_i.
    \label{lpsi}
\end{equation}
$\mathcal{Z}$ finally converges to:$\mathcal{Z}^*=(1-\alpha)(I-\alpha S)^{-1}Y$,where $S=D^{-1/2}AD^{-1/2}$ is the normalized weight matrix of graph $\mathcal{G}$, $\alpha$ is the fraction of label information from neighbors, and $\mathcal{N}(i)$ stands for the neighbor set of the node $i$. After obtaining the converged label matrix $\mathcal{Z}^*$, one node can be identified as a source when its final label is larger than its neighbors.
While node labels alone cannot fully capture complex structural information, this method still effectively identifies structural patterns related to source centrality~(Appendix~\ref{app:assumption}). In our work, we leverage the idea of LPSI to generate structural guidance for efficient source identification, and details can be found in Section 4.

\section{SIDSL: the Proposed Method}
\subsection{Diffusion Model for Source Localization}
To capture the indeterminacy of the ill-posed localization problem, it is essential to build a probabilistic model that can also leverage the topological information in the graph structure. We consider using the generative diffusion model (DM) framework to tackle this requirement by modifying it as a source predictor, which classifies nodes as source or non-source. The diffusion model aims to learn source distributions by gradually adding Gaussian noise in samples in the forward process and learning to reverse this process using denoising networks conditioned on observations and graph structure. 

The forward process of the diffusion model is to gradually corrupt the initial source labels~($X_0=X$) by adding Gaussian noise over $n$ timesteps, which can be formulated as:
\begin{equation}
    \begin{aligned}
        p(X_{1:n}|X_0,Y,\mathcal{G}) &= \prod_{t=1}^n p(X_t|X_{t-1},Y,\mathcal{G}),\\
        p(X_t|X_{t-1},Y,\mathcal{G}) &= \mathcal{N}(X_t; \sqrt{1-\beta_t}X_{t-1}, \beta_tI), 
    \end{aligned}
\end{equation}
where $\beta_t$ is the noise schedule, $X_0$ the initial source labels, $Y$ the observations, $\mathcal{G}$ the graph structure, and $X_t$ the noisy features at timestep $t$. At $t=n$, $X_t$ becomes pure Gaussian noise, while the reverse process leverages $Y$ and $\mathcal{G}$ to guide denoising.

In the reverse process, it gradually transforms pure Gaussian noise into the original source labels, generating samples $\hat{X}\in[0,1]^{|V|\times1}$ representing source probabilities, with final predictions obtained by thresholding. To enhance stability across different propagation patterns, we leverage graph structure-based source estimations as conditional priors. 
These priors $X_{est}=\mathcal{Z}^*$, standing for potential sources, are obtained through label propagation (Section \ref{subsec:lpsi}), aggregating infection values through the graph's normalized weight matrix to capture source centrality and structural importance.

The reverse process can be formulated as:
\begin{equation}
    \begin{aligned}
        q(X_{n-1:0}|X_n,Y,\mathcal{G}) &= \prod_{t=n}^1 q(X_{t-1}|X_t,Y,\mathcal{G}),\\
        q(X_{t-1}|X_t,Y,\mathcal{G}) &= \mathcal{N}(X_{t-1}; \mu_\theta(X_t,t,Y,\mathcal{G},X_{est}), \sigma_t^2I),
    \end{aligned}
\end{equation}
where $\mu_\theta$ is parameterized by a denoising network $f_\theta$ that predicts the mean of the Gaussian distribution and $\sigma_t$ is the predicted variance. 
The iterative diffusion process enables fine-grained prior knowledge integration through progressive denoising. By incorporating topology-aware priors $X_{est}$, our framework combines structural knowledge with data-driven learning, enhancing generalization across propagation scenarios.

\begin{figure}[t]
\centering
  \includegraphics[width=0.95\linewidth]{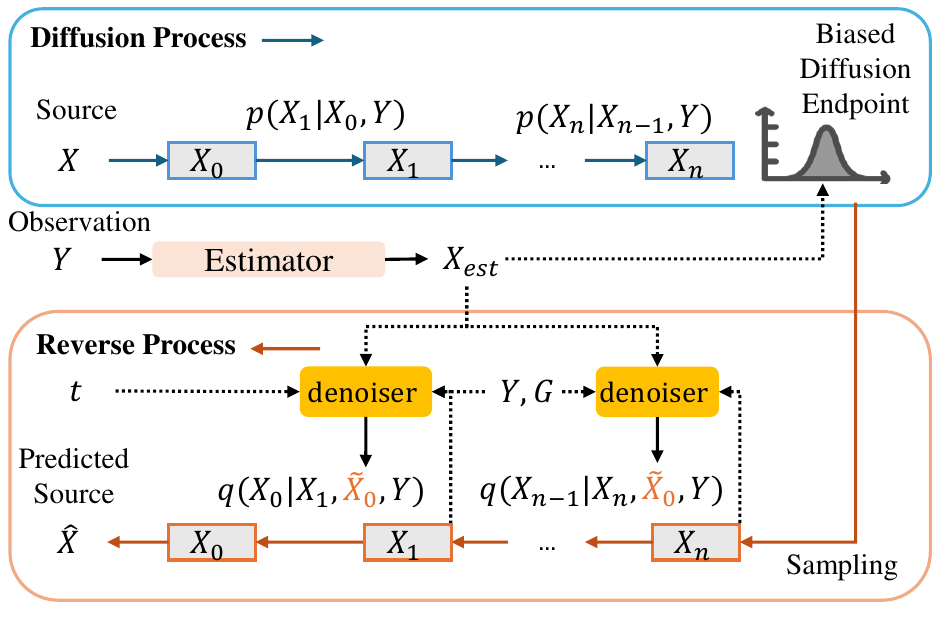} 
  \vspace{-1em}
  \caption{The framework of SIDSL.}
    \label{fig:framework}
    \vspace{-2em}
\end{figure}

\subsection{Structure-prior Biased Denoising Process}
Source nodes constitute a small minority of total graph nodes~\cite{cheng2024gin}. With limited training data, this severe class imbalance leads models to degenerate towards predicting all nodes as non-sources. Following~\citet{han2022card}, we propose a structure-prior biased denoising scheme that sets the prior to structure-based source estimations rather than random noise (Figure~\ref{fig:framework}). This leverages two insights: (1) structure-based estimations capture inherent source properties like centrality, providing more informative priors than random noise, and (2) initializing denoising from these estimations biases exploration toward regions with higher source likelihood. Formally, this approach decomposes the posterior as $p(X_0|Y,\mathcal{G}) \propto p_{{prior}}(X_{{est}}|Y,\mathcal{G}) \cdot p(X_0|X_{{est}})$, where the structural prior $p_{{prior}}$ anchors learning in high-probability source regions, enabling the denoiser $p(X_0|X_{{est}})$ to focus on lower-variance refinements that are learnable with scarce data.  This modified forward process shifts the diffusion trajectory from converging to noise toward structure-informed regions, while preserving generative correctness as our sampling process matches the modified forward posterior's functional form. These topology-aware priors prevent model collapse while reducing data requirements for accurate source learning.

Specifically, we first modify the mean of the diffusion endpoint (or reverse starting point) as the graph structure-based source estimation $X_{est}=X_{est}(Y,\mathcal{G})$ instead of using standard Gaussian noise, i.e.:
\begin{equation}
    p(X_n|Y,\mathcal{G})=\mathcal{N}(X_{est}(Y,\mathcal{G}),I).
\end{equation}
According to the original notation in \citet{ho2020denoising}, the Markov transition should be modified as:
\begin{equation}
    p(X_t|X_{t-1},Y,\mathcal{G})=\mathcal{N}(\sqrt{1-\beta_t}X_{t-1}+(1-\sqrt{1-\beta_t})X_{est},\beta_t I),
\end{equation}
which derives the closed-form distribution with arbitrary $t$:
\begin{equation}
    p(X_t|X_0,Y,\mathcal{G})=\mathcal{N}(\sqrt{\bar\alpha_t}X_{0}+(1-\sqrt{\bar\alpha_t})X_{est}(Y,\mathcal{G}),(1-\bar\alpha_t)I),
\end{equation}
where $\alpha_t:=1-\beta_t, \bar\alpha_t:=\prod_t\alpha_t$. 

In the reverse denoising process, the reverse Markov denoiser $q(X_{t-1}|X_t,Y,\mathcal{G})$ recovers the original data. DM framework trains the parameterized denoiser to fit the ground truth forward process posterior:
\begin{equation}
\begin{aligned}
    p(X_{t-1}|X_t,X_0,Y,\mathcal{G})&=
    p(X_{t-1}|X_t,X_0,X_{est}(Y,\mathcal{G}))\\
    &=\mathcal{N}(\tilde{\mu}(X_t,X_0,X_{est}(Y,\mathcal{G})),\tilde{\beta_t}I),
\end{aligned}
\end{equation}
where
\begin{equation}
  \begin{aligned}
  \tilde{\mu}(X_t,X_0,X_{est}(Y,\mathcal{G})):=&\frac{\sqrt{\bar\alpha_{t-1}}\beta_t}{1-\bar\alpha_t}X_0+\frac{(1-\bar\alpha_{t-1})\sqrt{\alpha_t}}{1-\bar\alpha_t}X_t+ \\
  &(1+\frac{(\sqrt{\bar\alpha_t}-1)(\sqrt{\alpha_t}+\sqrt{\bar\alpha_{t-1}})}{1-\bar\alpha_t})X_{est}, \\
  \tilde\beta_t:=&\frac{1-\bar\alpha_{t-1}}{1-\bar\alpha_t}\beta_t.
  \end{aligned}
\label{eq9}
\end{equation}
This formulation requires that the denoiser $f_\theta$ learns to predict $X_0$ given the modified trajectory, ensuring the sampling posterior $q_\theta(X_{t-1}|X_t,X_{est})$ matches the ground-truth form $p(X_{t-1}|X_t,X_0,X_{est})$ with $X_0$ replaced by the denoiser's prediction. The denoising network outputs the estimated source vector $\tilde{X}_0:=f_\theta(X_t,X_{est},Y,\mathcal{G},t)$ to calculate the posterior for step-by-step denoising. The denoising network $f_\theta$ can be trained by the simple L2 loss function~\cite{ho2020denoising}:
\begin{equation}
\label{eq:train}
    L(\theta)=\mathrm{E}_{X_0\sim p(X_0|\cdot),t,\epsilon}||X_0-f_\theta(X_t,t,\cdot)||^2_2,
\end{equation}
where $\cdot$ represents the conditional inputs.

\begin{figure}[t]
\centering
  \includegraphics[width=0.55\linewidth]{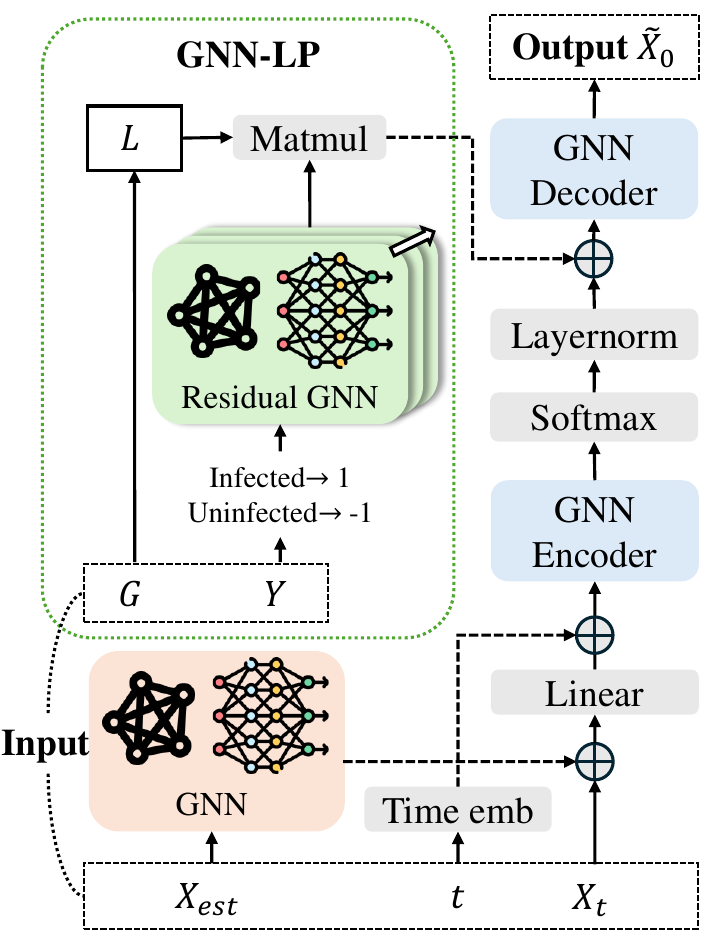} 
  \vspace{-1em}
  \caption{The architecture of the denoising network.}
    \label{fig:denoising}
    \vspace{-1em}
\end{figure}

\subsection{Propagtion-enhanced Conditional Denoiser}
\subsubsection{Denoising Network Architecture}

The architecture of our denoising network is shown in Figure~\ref{fig:denoising}. 

\textbf{Encoding the noisy input and soft labels.} The pre-estimated $X_{est}$ is forwarded through a multi-layer GNN to capture the hidden message with graph structural information. Subsequently, it is added to the noisy input $X_t$ and passed through a linear layer. The final input for the GNN encoder is $Z_{e}=\mathrm{Linear}(\mathrm{GNN}(X_{est})\oplus X_t)\oplus Emb(t),$ 
where for the denoising step t, we use the classical sinusoidal embedding~\citep{vaswani2017attention}. The $\oplus$ indicates element-wise sum.  
$Z_e$ is then passed through a GCN-based encoder and is smoothed through a softmax function $\sigma$ and layer normalization:
$Z_{d}=\mathrm{LN}(\sigma(\mathrm{GNN}(Z_e))).$

Softmax and layer normalization operations are then used to improve the network's representational capacity and convergence performance, resulting in better performance and faster training~\citep{huang2023normalization}.

\textbf{Conditioning.} Shown on the left part of the figure, a GCN-based module learns the encoding carrying the source prominence and centrality from the infection state input $Y$, which will be elaborated on in the next section.

\textbf{Decoder.} $Z_d$  and encoded condition $h_{out}$ are decoded through a GCN-based module, resulting in the estimation for the uncorrupted sample $X_0$ (i.e. $X$):
$\tilde{X_0} = \mathrm{GNN}(Z_d, h_{out}).$

\subsubsection{GNN-parameterized Label Propagation(GNN-LP)}
Our conditioning module takes the observed infection states as input, which contain crucial coupling information between spreading dynamics and network topology. To effectively extract this information, we introduce the GNN-LP module. We first employ label propagation~\citep{wang2017multiple}, which first marks infected nodes as 1 and uninfected as -1 in the observation state $Y$, resulting in $Y^*$. 
The propagation follows the update rule: the label of a node in the next step is a combination of its original label and the sum of normalized labels from its neighbors. We can rewrite this iteration as:
\begin{equation}
    \mathcal{Z}^{t+1}_i=\hat\alpha Y_i^*+\sigma(\sum_{j:j\in\mathcal{N}(I)}\phi(\mathcal{Z}_j^t, S_{ij})),
\end{equation}
where we add non-linear transformations $h(\cdot)$ and $\sigma(\cdot)$ to enhance the expressiveness of the propagation process.
During propagation, node labels evolve with the graph structure, reflecting node importance and the relationship between propagation patterns and network structure. We enhance this with GNNs, which learn topology-specific message passing rules to capture how local structures influence propagation. This method's formulation matches the general GNN~\citep{gilmer2017neural} framework and is parameterized by a GCN~\cite{kipf2016semi} combined with a residual block:
\begin{equation}
    \begin{aligned}
        g(h^{(l)})=\sigma(\tilde{D}^{-1/2}\tilde{A}\tilde{D}^{-1/2}\cdot h^{(l)}\cdot w),\\
        h^{(0)}=Y^*U^T,\quad h^{(l+1)}=h^{(0)}+ g(h^{(l)}).
    \end{aligned}
\end{equation}
Among them, $U\in\mathbb{R}^{C\times1}$ is the linear transformation, $\sigma$ is the activation operator {PReLU}, $h^{(l)}$ stands for the output hidden state of the $l$-th layer of the GCN, $\tilde{A}=A+I$ is the adjacency matrix with self-loops, and $\tilde{D}$ is the degree matrix of $\tilde{A}$. The final layer's output $h^{(l_f)}$ is projected back to dimension 1 and multiplied by the graph's Laplacian matrix $L$, i.e. $h_{out}:=L\cdot h^{(l_f)}$, which highlights the prominent nodes with higher propagation propagated labels among their neighbors. 

Through this integration, the module transforms binary infection states $Y$ into continuous representations $h_{out}$ that simultaneously encode both the spreading process and network structure information, enabling effective learning of their complex interactions even with limited training samples.

\begin{table*}[t]
\small
\renewcommand{\arraystretch}{0.85} 
\centering
\caption{Performance evaluation without pretraining under the real-world propagation. Best and second best (F1, RE, PR) are highlighted with bold and underlines respectively.}
\begin{tabular}{c|ccc|ccc|ccc|ccc}
\toprule
\multicolumn{1}{c|}{Datasets} & \multicolumn{3}{c|}{Digg} & \multicolumn{3}{c|}{Twitter} & \multicolumn{3}{c|}{Android} & \multicolumn{3}{c}{Christianity} \\
\cmidrule(lr){1-1} \cmidrule(lr){2-4} \cmidrule(lr){5-7} \cmidrule(lr){8-10} \cmidrule(lr){11-13}
Methods & F1 & RE & PR & F1 & RE & PR & F1 & RE & PR & F1 & RE & PR \\
\midrule
Netsleuth & 0.006 & 0.003 & 0.000 & 0.160 & 0.181 & 0.143 & 0.142 & 0.105 & 0.219 & 0.128 & 0.099 & 0.181 \\
LPSI & \underline{0.544} & 0.516 & \textbf{0.575} & 0.487 & 0.495 & 0.479 & 0.348 & 0.517 & 0.268 & 0.221 & 0.282 & 0.198 \\
GCNSI & 0.458 & 0.411 & 0.517 & 0.374 & 0.352 & 0.399 & 0.383 & 0.474 & 0.321 & 0.343 & 0.321 & 0.370 \\
TGASI & 0.472 & 0.406 & 0.564 & 0.362 & 0.327 & 0.405 & 0.388 & 0.462 & 0.335 & 0.377 & 0.339 & \underline{0.423} \\
SLVAE & 0.479 & 0.565 & 0.416 & 0.353 & 0.424 & 0.302 & \underline{0.467} & \underline{0.588} & 0.387 & \underline{0.458} & \underline{0.662} & 0.351 \\
DDMSL & 0.517 & \underline{0.592} & 0.459 & \underline{0.492} & \underline{0.504} & \underline{0.481} & 0.448 & 0.540 & \underline{0.432} & 0.417 & 0.481 & 0.368 \\
SIDSL(Ours) & \textbf{0.585} & \textbf{0.605} & \underline{0.566} & \textbf{0.546} & \textbf{0.516} & \textbf{0.580} & \textbf{0.522} & \textbf{0.702} & \textbf{0.439} & \textbf{0.519} & \textbf{0.747} & \textbf{0.436} \\
\midrule
$\Delta$ & +7.5\% & +2.2\% & -1.6\% & +11.0\% & +2.4\% & +20.6\% & +11.8\% & +48.3\% & +1.6\% & +13.3\% & +12.8\% & +3.1\% \\
\bottomrule
\end{tabular}
\label{tab:np_results}
\end{table*}

\begin{algorithm}[t]
\caption{Pretraining on synthetic data and few-shot learning on real data.}
\label{alg:pretrain}
\begin{algorithmic}[1]
\Require Graph $\mathcal{G}=(V,E)$, synthetic data generators $\mathcal{M}s$ (e.g. SIS, IC), few-shot samples $\mathcal{D}_{real}$
\State // Phase 1: Learn from Diverse Propagation Patterns
\State Generate synthetic $(X^*, Y)$ pairs using $\mathcal{M}s$ on $\mathcal{G}$
\State Train diffusion model using Eq.\ref{eq:train} with $(X^*, Y, \mathcal{G})$

\State // Phase 2: Adapt to Real Scenarios
\For{each $(X^*, Y)$ in $\mathcal{D}_{real}$}
    \State Fine-tune model using Eq.\ref{eq:train} with $(X^*, Y, \mathcal{G})$
\EndFor

\State // Inference on Real New Cases
\State \textbf{Input}: Observation $Y_{new}$ on $\mathcal{G}$
\State Compute structure-based prior $X_{est}$ from $Y_{new}$ using Eq.\ref{lpsi}
\State Sample initial noise $X_n \sim \mathcal{N}(0,I)$
\State \Return $\hat{X} \leftarrow$ denoise($X_n$) with $(X_{est}, Y_{new}, \mathcal{G})$

\end{algorithmic}
\vspace{-0.4em}
\end{algorithm}

\subsection{Pretrain Using Simulation Data}
\label{sec:method_pretrain}
The proposed structure-prior informed diffusion framework effectively extracts topology-aware features that remain stable across different propagation dynamics. This enables pretraining on synthetic data from established models (IC, LT, SIS) with efficient few-shot adaptation to real scenarios, addressing data-limited source localization. We illustrate this process in Algorithm~\ref{alg:pretrain} and evaluate SIDSL's simulation-to-reality transfer in Section~\ref{sec:exp_pre} by testing two approaches: one using pretrained models with few-shot learning, and another trained exclusively on real-world data.

\begin{table*}[ht]
\small
\renewcommand{\arraystretch}{0.85} 
\centering
\caption{Few-shot learning performance evaluation under the real-world propagation with pretraining (P) and without pretraining (NP). Results show performance using simulation data(IC and LT) for pretraining, and using limited real-world data (10\%) for few-shot learning. The rule-based methods are omitted as they do not support pretraining. Best and second best are highlighted with bold and underline respectively.}
\begin{tabular}{c|c|ccc|ccc|ccc|ccc} 
\toprule 
\multicolumn{2}{c|}{Dataset} & \multicolumn{3}{c|}{Digg} & \multicolumn{3}{c|}{Twitter} & \multicolumn{3}{c|}{Android} & \multicolumn{3}{c}{Christianity} \\ 
\cmidrule(lr){1-2} \cmidrule(lr){3-5} \cmidrule(lr){6-8} \cmidrule(lr){9-11} \cmidrule(lr){12-14} 
Methods & Config & F1 & RE & PR & F1 & RE & PR & F1 & RE & PR & F1 & RE & PR \\ 
\midrule 
\multirow{2}{*}{GCNSI} & P & 0.382 & 0.398 & 0.367 & 0.291 & 0.251 & 0.346 & 0.194 & 0.145 & \underline{0.293} & 0.159 & 0.174 & \underline{0.146} \\ 
& NP & 0.162 & 0.201 & 0.136 & 0.003 & 0.015 & 0.002 & \underline{0.012} & 0.029 & 0.008 & \underline{0.176} & \underline{0.179} & 0.178 \\ 
\midrule 
\multirow{2}{*}{TGASI} & P & \underline{0.406} & 0.407 & \underline{0.405} & \underline{0.292} & 0.249 & \underline{0.353} & \underline{0.212} & \underline{0.184} & 0.251 & \underline{0.206} & 0.193 & \underline{0.221} \\ 
& NP & \underline{0.297} & \underline{0.303} & \underline{0.291} & 0.003 & 0.015 & 0.001 & 0.012 & \underline{0.062} & 0.006 & 0.036 & 0.038 & 0.035 \\ 
\midrule 
\multirow{2}{*}{SLVAE} & P & 0.315 & 0.441 & 0.246 & 0.195 & 0.325 & 0.139 & 0.036 & 0.132 & 0.021 & 0.171 & \underline{0.215} & 0.142 \\ 
& NP & 0.103 & 0.188 & 0.071 & 0.030 & 0.042 & 0.023 & 0.009 & 0.010 & 0.008 & 0.008 & 0.016 & 0.005 \\ 
\midrule 
\multirow{2}{*}{DDMSL} & P & 0.348 & \underline{0.504} & 0.266 & 0.250 & \underline{0.359} & 0.192 & 0.109 & 0.114 & 0.104 & 0.165 & 0.145 & 0.191 \\ 
& NP & 0.005 & 0.003 & 0.015 & \underline{0.101} & \underline{0.095} & \underline{0.108} & 0.010 & 0.015 & \underline{0.009} & 0.009 & 0.005 & 0.045 \\ 
\midrule 
\multirow{2}{*}{SIDSL(Ours)} & P & \textbf{0.571} & \textbf{0.623} & \textbf{0.527} & \textbf{0.518} & \textbf{0.503} & \textbf{0.547} & \textbf{0.547} & \textbf{0.567} & \textbf{0.529} & \textbf{0.483} & \textbf{1.000} & \textbf{0.319} \\ 
& NP & \textbf{0.384} & \textbf{0.409} & \textbf{0.362} & \textbf{0.120} & \textbf{0.118} & \textbf{0.143} & \textbf{0.018} & \textbf{0.253} & \textbf{0.015} & \textbf{0.432} & \textbf{0.757} & \textbf{0.309} \\ 
\midrule
\multirow{2}{*}{$\Delta$} & P & +41\% & +24\% & +30\% & +77\% & +40\% & +55\% & +158\% & +208\% & +81\% & +134\% & +365\% & +44\% \\
& NP & +29\% & +35\% & +24\% & +19\% & +24\% & +32\% & +50\% & +308\% & +67\% & +145\% & +323\% & +74\% \\
\bottomrule 
\end{tabular}
\label{tab:few_shot}
\vspace{-1em}
\end{table*}

\section{Experiments}
\subsection{Experiment Settings}
The basic settings of our experiments are shown below.

\textbf{Datasets.} Datasets. Following~\citet{ling2022source} and~\citet{huang2023two}, we evaluate SIDSL on four real-world propagation datasets containing actual cascades: \textit{Digg}, \textit{Twitter}, \textit{Android}, and \textit{Christianity}. For each cascade, we define source nodes as those infected within the first 10\% of the propagation time and use the network's infection status at 30\% of the time as the observation input. This setup simulates real-world applications, where initial events must be inferred from data collected after a delay. Further dataset details are in Appendix~\ref{app:data}.

\textbf{Implementation Details.}
We use a 6:1:1 train/validation/test split for all datasets. The diffusion framework employs $T=500$ timesteps with linear noise scheduling. The denoising network uses a 2-layer GCN for LPSI estimation $X_{est}$ conditioning, while the GNN encoder/decoder comprises 3-layer GCNs (hidden dimension 128) and the residual GNN uses 2 layers (hidden dimension 8). We use Adam optimizer with learning rate searched from {0.01, 0.005, 0.001}, linear decay scheduling, and maximum 500 training epochs.


\textbf{Baselines.} Following previous works~\citep{ling2022source, yan2024diffusion}, we selected two representative heuristic methods, \textit{i.e.}, Netsleuth~\citep{prakash2012spotting} and LPSI~\citep{wang2017multiple}, and deep learning methods, \textit{i.e.}, GCNSI~\citep{dong2019multiple}, TGASI~\citep{hou2023sequential} and recent generative deep learning methods SLVAE~\citep{ling2022source}, DDMSL~\citep{yan2024diffusion} that capture source localization uncertainty. These baselines are all state-of-the-art (SOTA) multi-source localization methods in their domains. Please refer to the Appendix~\ref{app:baseline} for specific information on baselines and our method.

\textbf{Metrics.} Following~\citep{wang2022invertible}, we use three metrics: 1) F1-score (F1): The harmonic mean of precision and recall; 2) Recall (RE): The proportion of correctly identified positive cases (source nodes); and 3) Precision (PR): The proportion of true positives among predicted positive cases. We omit accuracy because it is a misleading metric in highly imbalanced datasets, as it fails to capture the model's ability to identify the rare positive class.

\subsection{Overall Performance on Real-world Datasets}

To evaluate the real-world performance of our proposed method against baselines, we conduct direct training and testing on four real-world datasets for all methods. 
The experimental results are presented in Table \ref{tab:np_results}. Our proposed SIDSL outperforms all baselines across nearly all metrics on all datasets. Specifically, SIDSL achieves F1 scores that exceed the second-best baseline by 7.5\%, 11.0\%, 11.8\%, and 13.3\% on Digg, Twitter, Android, and Christianity, respectively, which demonstrates SIDSL's superior ability to predict source nodes despite their sparsity. Our model also excels in recall, with significant gains in Android (48.3\%) and Christianity (12.8\%), demonstrating strong ability in identifying true source nodes. 

Further, we have the following findings. \textbf{First}, compared to purely data-driven generative methods like DDMSL and SLVAE, SIDSL demonstrates superior performance, validating the effectiveness of incorporating structural prior information.
\textbf{Second}, compared to rule-based and learning-based methods, generative methods (SIDSL, SLVAE, DDMSL) show stronger performance, highlighting the advantages of effective source distribution modeling. LPSI and other deep learning-based methods perform comparably well, successfully capturing structural characteristics. Netsleuth demonstrates the weakest performance due to limited generalization beyond SI propagation patterns~\cite{keeling2005networks,prakash2012spotting}.
\textbf{Finally}, SIDSL shows the most substantial improvement on the Christianity dataset (13.3\% in F1), as a smaller network scale enhances topological feature capture, amplifying structural prior benefits. 

\begin{table*}[ht]
\small
\renewcommand{\arraystretch}{0.9} 
\setlength{\tabcolsep}{5pt}
\centering
\caption{Zero-shot performance evaluation with pretraining under the real-world propagation. Results show performance without finetuning when using simulation data(IC and LT) for pretraining. The rule-based methods are omitted as they do not support pretraining. Best and second best are highlighted with bold and underline respectively.}
\begin{tabular}{c|ccc|ccc|ccc|ccc}
\toprule
\multicolumn{1}{c}{Dataset} & \multicolumn{3}{|c}{Digg} & \multicolumn{3}{|c}{Twitter} & \multicolumn{3}{|c}{Android} & \multicolumn{3}{|c}{Christianity} \\
\cmidrule(lr){1-1} \cmidrule(lr){2-4} \cmidrule(lr){5-7} \cmidrule(lr){8-10} \cmidrule(lr){11-13}
Methods & F1 & RE & PR & F1 & RE & PR & F1 & RE & PR & F1 & RE & PR \\
\midrule
GCNSI & 0.195 & 0.273 & 0.152 & \underline{0.203} & 0.162 & \underline{0.271} & 0.103 & 0.079 & \underline{0.148} & 0.030 & 0.060 & 0.020 \\
TGASI & 0.189 & 0.281 & 0.142 & 0.199 & 0.172 & 0.236 & \underline{0.109} & \underline{0.107} & 0.110 & 0.059 & 0.067 & 0.052 \\
SLVAE & 0.278 & 0.236 & \underline{0.337} & 0.172 & 0.205 & 0.148 & 0.004 & 0.025 & 0.002 & 0.080 & \underline{0.140} & 0.056 \\
DDMSL & \underline{0.291} & \underline{0.339} & 0.264 & 0.203 & \underline{0.272} & 0.162 & 0.019 & 0.013 & 0.038 & \underline{0.133} & 0.127 & \underline{0.147} \\
SIDSL(Ours) & \textbf{0.539} & \textbf{0.629} & \textbf{0.472} & \textbf{0.509} & \textbf{0.569} & \textbf{0.460} & \textbf{0.506} & \textbf{0.914} & \textbf{0.355} & \textbf{0.465} & \textbf{1.000} & \textbf{0.305} \\
\midrule
$\Delta$ & +85\% & +86\% & +40\% & +151\% & +109\% & +70\% & +364\% & +754\% & +140\% & +250\% & +614\% & +108\% \\
\bottomrule
\end{tabular}
\label{tab:zero_shot}
\vspace{-1em}
\end{table*}

\subsection{Performance in Low Data Regime}
\label{sec:exp_pre}

To assess how deep learning methods leverage simulation data pretraining for few-shot and zero-shot learning in real-world source localization, we generate pretraining data using standard IC and LT propagation models (1:1 ratio) across four networks. These models effectively capture social network dynamics by representing real-world interaction randomness, cumulative influence, and topology-based spread patterns. For few-shot learning~(P), we randomly selected 10\% of the training data, sharing this data across all evaluated methods. We also compare the performance of all the methods trained on equivalent data volumes but without pretraining~(NP). For zero-shot learning, all the methods are directly tested after being pretrained. 

\textbf{Few-shot analysis.} The result of few-shot learning is shown in Tabel~\ref{tab:few_shot}. Rule-based methods (Netsleuth and LPSI) are omitted as they are training-free. SIDSL demonstrates superior performance compared to all baseline methods across all datasets, both with and without pretraining, achieving improvements of 24-365\% across all metrics.
We observe two key findings. \textbf{First}, SIDSL demonstrates enhanced performance advantages when leveraging pretraining, with 8 out of 12 cases showing larger gains with P compared to NP across different datasets and metrics. This pattern of larger improvements with pretraining demonstrates SIDSL's superior ability to effectively leverage information from simulated data to real-world scenarios in limited-data settings.
\textbf{Second}, baseline methods exhibit varying capabilities in utilizing pretraining. While TGASI shows consistent improvements with pretraining, generative approaches (SLVAE, DDMSL) demonstrate limited gains (e.g., F1 increases from 0.009 to 0.036 for SLVAE, 0.010 to 0.109 for DDMSL on Android). This highlights generative models' sensitivity to distribution shifts, whereas our method achieves robust improvements through effective incorporation of pattern-invariant structural priors.

\textbf{Zero-shot analysis.} The results of zero-shot learning are shown in Table~\ref{tab:zero_shot}. SIDSL significantly outperforms all baseline methods across all datasets, achieving improvements up to 364\% in F1 scores. SIDSL achieves particularly strong performance on Android and Christianity datasets (>0.9 recall with reasonable precision). Baseline methods show limited zero-shot capability (F1 < 0.3). Zero-shot SIDSL outperforms rule-based methods (Table~\ref{tab:np_results}) in most metrics. These findings demonstrate SIDSL's superior simulation-to-real transfer without fine-tuning.

In few-shot and zero-shot settings, pretrained SIDSL achieves higher recall than precision gains, effectively identifying true sources but with occasional false positives, especially in Android and Christianity datasets. This suggests strong propagation modeling despite some errors. For misinformation tracing, pretrained SIDSL provides superior source identification. Accuracy can be enhanced using precision-based classifiers with boosting to filter false positives. Without pretraining, SIDSL outperforms other methods on both metrics, proving effective in low-data scenarios.

\begin{figure}[t]
\centering
  \includegraphics[width=0.43\linewidth]{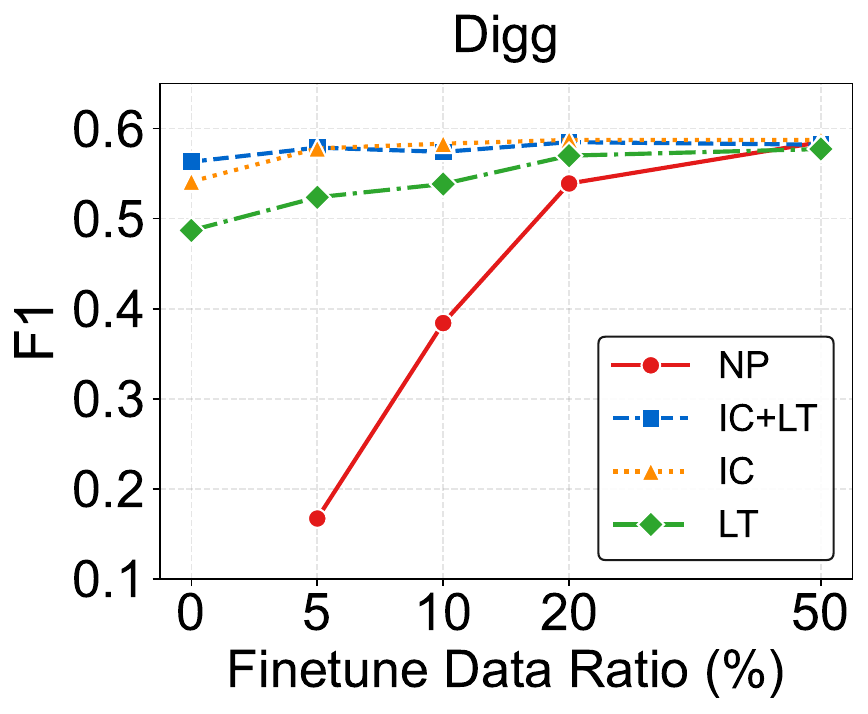} 
  \includegraphics[width=0.43\linewidth]{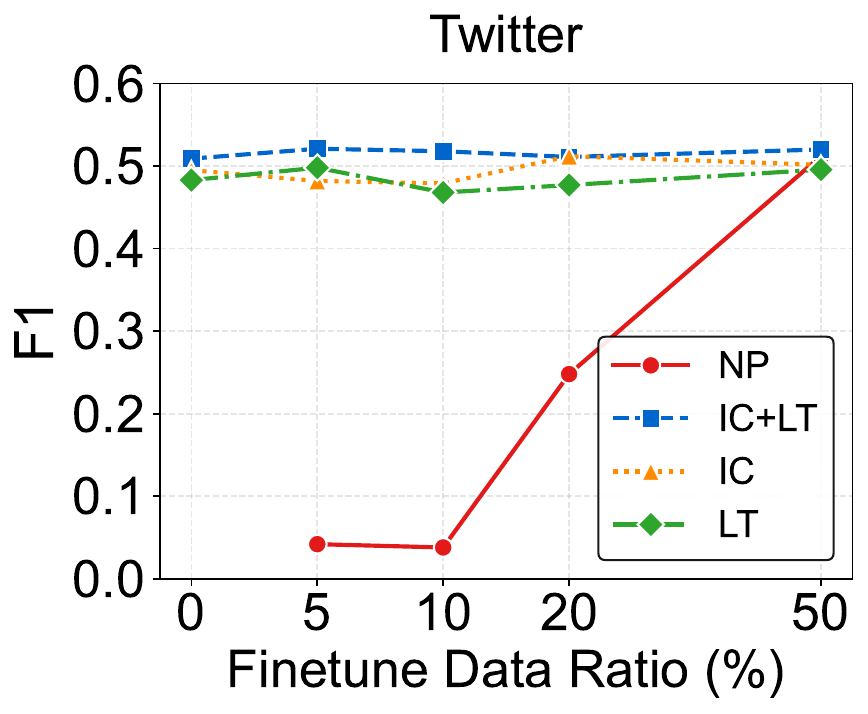} 
  \includegraphics[width=0.43\linewidth]{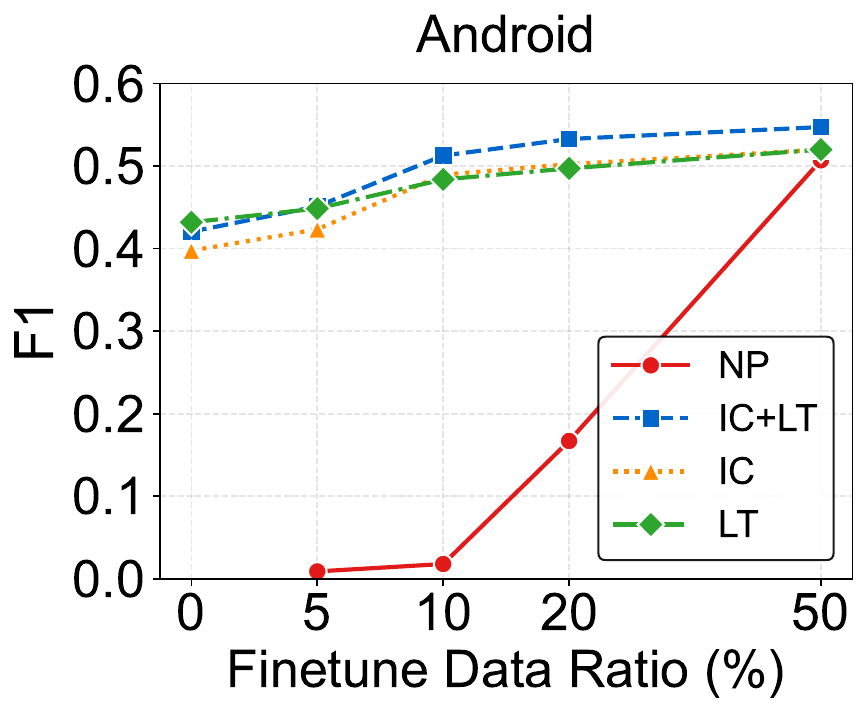} 
  \includegraphics[width=0.43\linewidth]{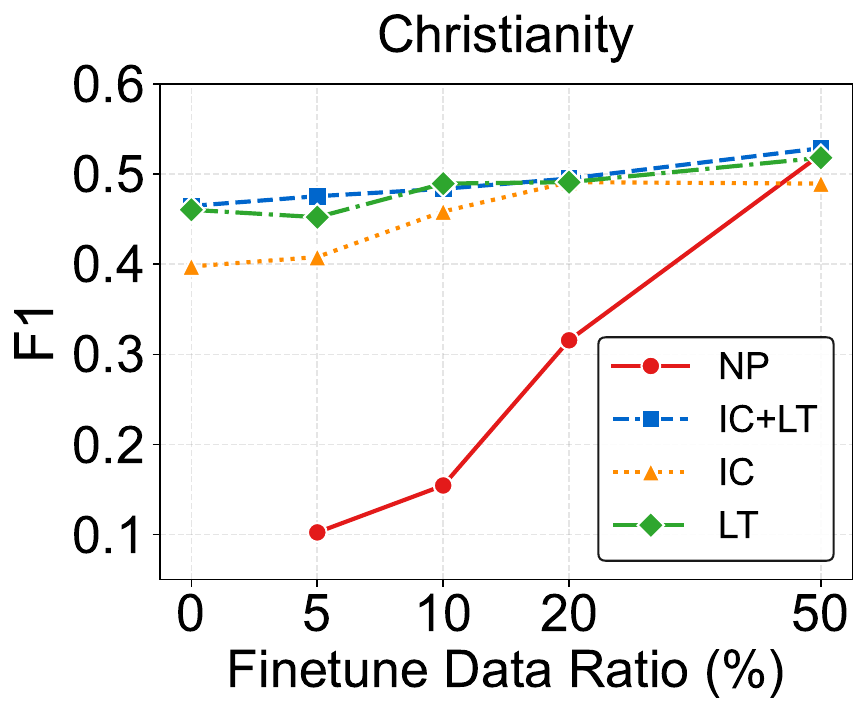} 
  
  \caption{Few-shot performance of SIDSL with different pretraining dataset compositions~(IC+LT, IC, LT), then finetuned on real-world data. "NP" denotes "non-pretrained".}
\label{fig:pattern}
\vspace{-1em}
\end{figure}

\subsection{Analysis of Pretraining on Simulation Data}
To investigate how pretraining on simulation data affects the model's few-shot/zero-shot performance under different data volumes, we evaluate SIDSL's F1 scores across the pretraining dataset with different propagation model combinations (IC, LT, and IC+LT with a ratio of 1:1) and compare them with non-pretrained~(NP) SIDSL. We then test them under different ratios of the finetune dataset of real-world propagation data. The results of four datasets are shown in Figure~\ref{fig:pattern}.

\textbf{Finetune data volume analysis.}  Across all four datasets, models with pretraining consistently outperformed non-pretrained (NP) models in low-data scenarios. Under the IC+LT combination, the pretrained models achieved optimal F1 scores on Digg and Twitter using just 5\% of training data, while NP models required 50\%. Similarly, on Android and Christianity datasets, pretrained models reached near-peak performance (within 0.033 and 0.018 F1) with only 20\% of the data, compared to NP models, which needed 50\%. These findings suggest that aligning pretraining patterns with specific propagation characteristics benefits model performance.

\textbf{Pattern analysis.} On the Digg dataset, SIDSL pretrained with IC+LT and IC configurations outperform those using LT, indicating that IC patterns better align with Digg's propagation characteristics, as also shown in~\cite{zhang2024information}. For Android and Christianity Q\&A datasets, IC+LT and LT configurations performed better in few-shot scenarios. This is likely because user participation in these platforms' information cascades (e.g., comments) is driven by cumulative influence from multiple sources (LT) rather than single independent triggers (IC). 
On Twitter, the IC+LT configuration performs the best, demonstrating that pattern diversity enhances pretraining effectiveness on this dataset.

\begin{table}
\footnotesize
\setlength{\tabcolsep}{4pt}
\centering
\caption{Ablation study. The relative performance change of each ablation against SIDSL is reported.}
\vspace{-1.5em}
\begin{tabular}{llcccccc}
\toprule
Dataset & & \multicolumn{3}{c}{Digg} & \multicolumn{3}{c}{Christianity} \\
\cmidrule(lr){1-8}
Data & Ablation & F1 & RE & PR & F1 & RE & PR \\
\cmidrule(lr){1-2} \cmidrule(lr){3-5} \cmidrule(lr){6-8}
\multirow{3}{*}{100\%} & SIDSL & 100\% & 100\% & 100\% & 100\% & 100\% & 100\% \\
& SIDSL w/o SBD & -34\% & -47\% & -16\% & -19\% & -49\% & +2\% \\
& SIDSL w/o GNN-LP & -7\% & -9\% & -5\% & -4\% & -6\% & -5\% \\
\midrule
\multirow{3}{*}{10\%} & SIDSL & 100\% & 100\% & 100\% & 100\% & 100\% & 100\% \\
& SIDSL w/o SBD & -46\% & -54\% & -36\% & -35\% & -55\% & -24\% \\
& SIDSL w/o GNN-LP & -17\% & -16\% & -17\% & -21\% & -34\% & -17\% \\
\bottomrule
\end{tabular}
\label{tab:ablation}
\vspace{-1em}
\end{table}

\subsection{Ablation Study}
We conduct ablation studies to investigate the significance of key components in SIDSL. For the first ablation model, SIDSL w/o SBD (structure-prior biased diffusion), instead of computing estimation through Equation~\ref{lpsi}, we directly substitute it with the node having the *highest closeness centrality* in the infection subgraph. For the second ablation, SIDSL w/o GNN-LP, we replace the observation state encoding with a standalone GNN while maintaining an equivalent parameter count. 
Table~\ref{tab:ablation} presents the ablation results on Digg and Christianity datasets under both full (100\%) and limited (10\%) training data conditions. Removing either component leads to decreased overall performance under both conditions. There are two more findings:
\textbf{First}, removing SBD leads to more severe performance degradation compared to removing GNN-LP, with the impact most pronounced in recall metrics. The substantial decline in recall (TP/(TP+FN)) indicates reduced true positive predictions. The decline in recall exceeds that of precision (TP/(TP+FP)), implying fewer positive predictions overall~(TP+FP). This demonstrates SBD's role in maintaining predictive balance by ensuring adequate positive predictions.
\textbf{Second}, the impact of removing GNN-LP shows greater impact under limited data (F1 drops: -17\% to -21\% at 10\% vs. -4\% to -7\% at full data), demonstrating GNN-LP's effectiveness in low-resource scenarios.


\begin{figure}[t]
\centering
  \includegraphics[width=0.9\linewidth]{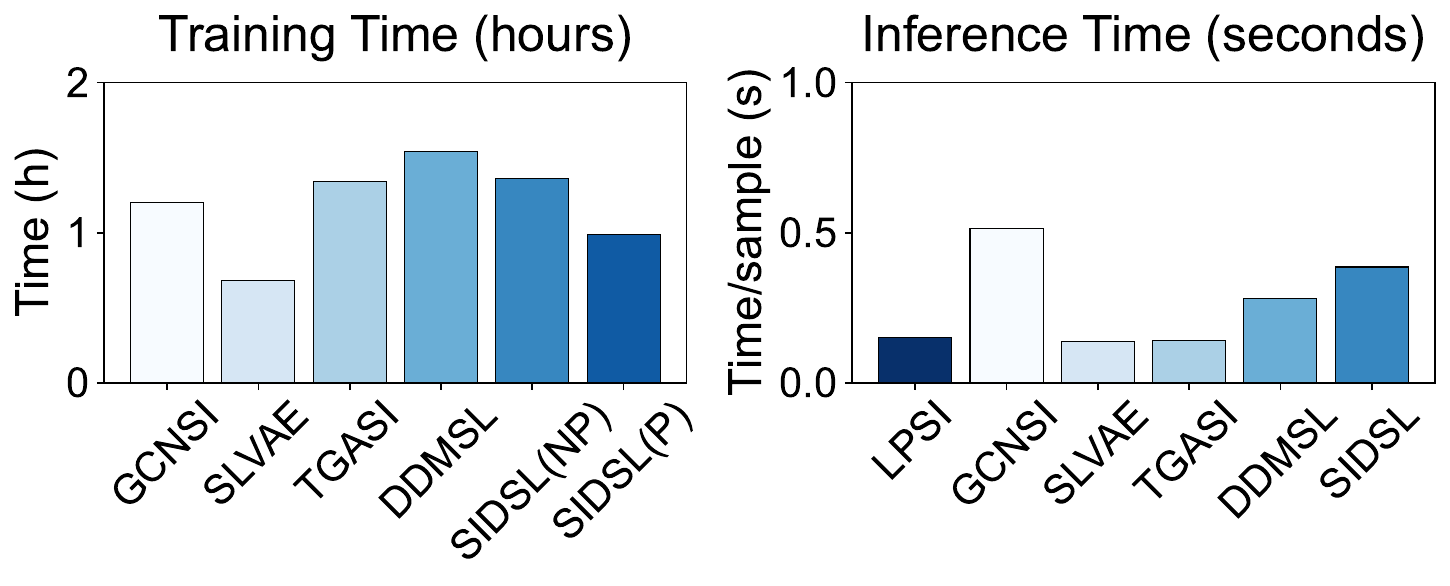} 
  \vspace{-1em}
  \caption{Training and inference time~(per sample) comparison on Digg dataset. "NP" denotes the "non-pretrained" version. "P" denotes using the scheme of "pretraining+finetuning", and the overall training time is reported.}
    \label{fig:time}
    \vspace{-1.5em}
\end{figure}

\subsection{Time Cost Analysis}
We present a detailed comparison of the computational cost of our proposed model against baselines in the Digg dataset, which has one of the largest networks for better comparison. Results are shown in Figure \ref{fig:time}. While our model's original training duration is the second highest, our model requires less time to pretrain on the synthetic data and finetune than initially training on the real data~(-25\%), which makes the training time the second lowest. 

In inference time, our model is the second-highest because of the iterative nature of the DDPM-based denoising process. We opt for DDPM as our foundation due to its classical design and proven effectiveness. Note that SIDSL's architecture is fully compatible with more computationally efficient diffusion variants, such as DDIM~\cite{song2020denoising}, which could substantially reduce the current computational overhead. This flexibility, combined with our model's transfer capabilities, makes SIDSL particularly resource-efficient in practical deployments. 

\begin{figure}[t]
\centering
  \includegraphics[width=0.99\linewidth]{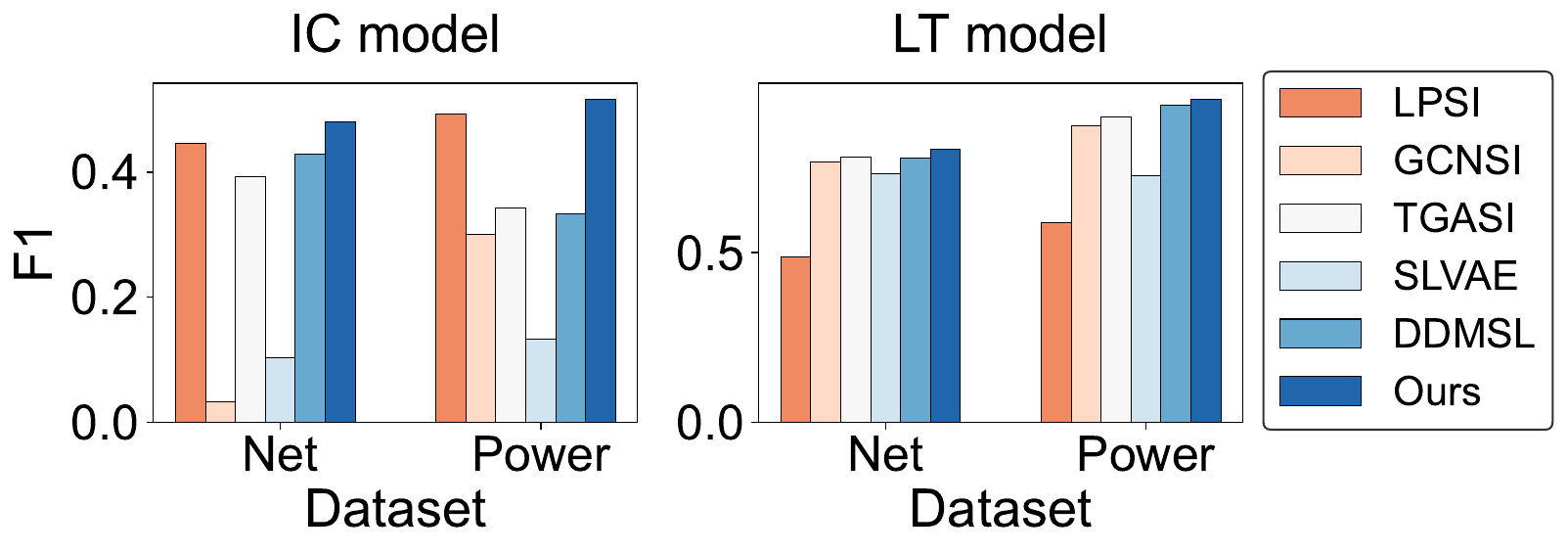} 
  \vspace{-1em}
  \caption{Performance under IC and LT propagation patterns.}
    \label{fig:iclt_simulation}
\vspace{-1em}
\end{figure}

\subsection{Results on Synthetic Data}
It is a common practice in previous works~\cite{ling2022source,yan2024diffusion,wang2022invertible} to evaluate methods' performance on synthetic data with established propagation patterns. Following \cite{ling2022source,wang2022invertible}, we conducted experiments on two commonly evaluated networks: \textit{Net Science~(Net)} and \textit{Power Grid(Power)}. Using both IC and LT models, we simulated 100 steps until convergence to generate two sets of synthetic data. Strong baseline models were trained and tested on these datasets, with results shown in Figure~\ref{fig:iclt_simulation}. Our method outperforms the baselines, achieving F1 score improvements of 7.6\% and 4.7\% under the IC model, and 2.7\% and 1.7\% under the LT model on the two datasets, respectively. These results demonstrate our method's ability to accurately identify sources with established propagation patterns.

\section{Conclusion and Future Work}
We introduced SIDSL, a structure-aware diffusion model that significantly enhances source localization in sparse graphs by integrating structural priors. Our framework demonstrates a notable improvement in F1 scores by 7.5-13.3\% over leading-edge methods, proving especially robust in few-shot and zero-shot scenarios where it maintains high performance with as little as 10\% of training data. This highlights its strong generalization capabilities and practical value in data-scarce environments.
However, the iterative nature of the diffusion process introduces computational overhead, which may limit its use in real-time applications. While the current architecture is not suited for million-node networks, its core principles can be adapted for hierarchical frameworks to analyze large-scale graphs. Future work will focus on improving computational efficiency and extending the framework to handle a wider variety of source types. These limitations also open up avenues for further research to enhance the model's scalability and robustness.


\appendix
\section*{Appendix}

\begin{table}[h]
\small
\centering
\caption{Overview of networks. CC: clustering coefficient.}
\vspace{-1em}
\setlength{\tabcolsep}{7pt}
\begin{tabular}{lccccc}
\hline
\textbf{Dataset} & Nodes & Edges  & Mean Degree & CC \\ \hline
\textbf{Digg}    & 14511 & 194405 & 13.39       & 0.1353                 \\
\textbf{Twitter}    & 12619 & 309621 & 24.52       & 0.2962                \\
\textbf{Android}    & 9958 & 42915 & 8.62       & 0.4121                \\
\textbf{Christianity}    & 2897 & 30044 & 20.74       & 0.6027                \\
\textbf{Jazz}    & 198   & 2742   & 13.84       & 0.6174                 \\
\textbf{Net Science}     & 1589  & 2742   & 1.72        & 0.6377                 \\
\textbf{Power Grid}   & 4941  & 6594   & 1.33        & 0.0801                 \\
\hline
\end{tabular}%
\label{tab:dataset}
\vspace{-1em}
\end{table}

\section{Dataset Description}\label{app:data}

The study utilizes both real-world and synthetically generated propagation data. This section details the datasets employed in our study, with network statistics summarized in Table~\ref{tab:dataset}.

The cascades in real-world propagation datasets represent information diffusion. These include Digg\citep{rossi2015network}, with cascades formed by votes on stories; Twitter\citep{yang2021full}, where each tweet's spread is a cascade; and the Stack Exchange communities Android and Christianity~\citep{huang2023two}, where cascades are chronologically ordered posts with identical tags. The top 10\% of nodes ranked by influence time were designated as diffusion sources for each cascade, with the top 30\% considered as observed influenced nodes.

Synthetic propagation datasets were generated using SIR, IC, and LT models on three real-world networks: Jazz\citep{rossi2015network}, Net Science\citep{rossi2015network}, and Power Grid~\citep{watts1998collective}.
The SIR model used constant infection and recovery rates. For IC and LT models, edge propagation probabilities were inversely proportional to the target node's degree. Simulations ran for 100 steps or until convergence.
The Jazz network represents musician collaborations; 5\% of nodes were randomly selected as sources.
The Net Science network is a coauthorship graph of network scientists,and the Power Grid network maps the Western US electrical grid topology; 0.5\% of nodes were randomly selected as sources in both datasets.

\begin{table}[t]
\footnotesize
\centering
\caption{Comparison of localization methods.\textbf{Obs.}: Observation input (single/multiple snapshots); \textbf{PI}: Prior-knowledge informed; \textbf{Patterns}: Applicable diffusion patterns.}
\setlength{\tabcolsep}{6pt}

\vspace{-1em}
\label{tab:comp}
\begin{tabular}{@{}llccll@{}}
\toprule
\textbf{Category} & \textbf{Method} & \textbf{Indeterminacy} & \textbf{Patterns} & \textbf{Obs.} & \textbf{PI} \\ \midrule
Rule-based & NetSleuth \citep{prakash2012spotting} & \ding{55} & SI & single & - \\ 
 & OJC \citep{zhu2017catch} & \ding{55} & SI, SIR & single & - \\ 
 & LPSI \citep{wang2017multiple} & \ding{55} & any & single & - \\ 
Data-driven & GCNSI \citep{dong2019multiple} & \ding{55} & any & single & \checkmark \\ 
 & IVGD \citep{wang2022invertible} & \ding{55} & IC & single & \ding{55} \\ 
 & SLVAE \citep{ling2022source} & \checkmark & any & single & \ding{55} \\ 
 & SLDiff \citep{huang2023two} & \ding{55} & any & multiple & \ding{55} \\ 
 & TGASI \citep{hou2023sequential} & \ding{55} & any & multiple & \ding{55} \\ 
 & DDMSL \citep{yan2024diffusion} & \checkmark & any & multiple & \checkmark \\ 
 & PGSL \citep{xu2024pgsl} & \checkmark & any & single & \ding{55} \\ 
 & GINSD \citep{cheng2024gin} & \ding{55} & any & single & \ding{55} \\ 
 & NFSL \citep{hou2024new} & \ding{55} & any & single & \checkmark \\ 
  & CRSLL \citep{hougood} & \ding{55} & any & single & \checkmark \\
 & SIDSL (Ours) & \checkmark & any & single & \checkmark \\ \bottomrule
\end{tabular}
\vspace{-1em}

\end{table}

\section{Comparison of Multiple Source Localization Methods}\label{app:comp}
Table~\ref{tab:comp} compares mainstream source localization methods on functionality, requirements, and applications. Our method is more functional, versatile, requires less input, and addresses existing limitations, offering significant practical value. Unlike other diffusion-based methods (DDMSL, TGASI), ours doesn't require propagation data or model parameters pre-localization. PGSL, using a flow-model, is surpassed by our diffusion model's superior distribution modeling. GINSD addresses incomplete data (not our focus), otherwise reducing to a GAT-based baseline like GCNSI.
Cross-platform works~\citep{wang2024joint,ling2024source} are orthogonal and not discussed.

\section{Baselines} \label{app:baseline}
We compare SIDSL against state-of-the-art baselines that locate sources from propagation snapshots without requiring knowledge of the underlying propagation pattern.

\textbf{NetSleuth}~\citep{prakash2012spotting} employs a Minimum Description Length approach for multiple sources, specifically within the Susceptible-Infected (SI) model.
\textbf{LPSI}~\citep{wang2017multiple} detects multiple sources using source prominence and label propagation; its parameter $\alpha$ was tuned from \{0.1, 0.3, 0.5, 0.7, 0.9\} for each dataset.
\textbf{GCNSI}~\citep{dong2019multiple} utilizes Graph Convolutional Networks for multiple rumor source identification, with settings as per the original paper.
\textbf{SLVAE}~\citep{ling2022source} is a probabilistic Variational Autoencoder for source localization that quantifies uncertainty; we used the original implementation and tuned the learning rate (0.001-0.05).
\textbf{TGASI}~\citep{hou2023sequential} offers a sequence-to-sequence framework (GNN encoder, GRU decoder with temporal attention) for multiple rumor source detection, considering heterogeneous user behavior and time-variance, designed for transferability with a unique loss function.
\textbf{DDMSL}~\citep{yan2024diffusion} proposes a probabilistic model for source localization and diffusion path reconstruction via a discrete denoising-diffusion model with a reversible residual network.

\section{Experiments and Implementation Details}
We partitioned datasets using a 6:1:1 train/validation/test split. SIDSL employed T=500 diffusion timesteps with a linear noise schedule, without classifier-free sampling. Grid search was conducted over encoder/decoder layers \{2, 4, 6\} and hidden dimensions \{32, 64, 128, 256\}. After comparing GNN variants (GCN, GIN, GAT, MLP), we selected a 2-layer GCN for LPSI estimation ($X_{est}$) as the denoising network. The encoder/decoder architecture consisted of 3-layer GCNs with 128-dimensional hidden layers, while the conditioning module employed a 2-layer GCN with 8-dimensional hidden layers. Models were trained using Adam optimizer with learning rates selected from \{0.01, 0.005, 0.001\} and linear decay, running for 500 epochs (300 for few-shot scenarios) on an RTX 2080 Ti.
\section{Analyzing Source Centrality in Empirical Data}
\label{app:assumption}
\begin{figure}[h]
    \centering
\vspace{-1em}
    
    \includegraphics[width=\linewidth]{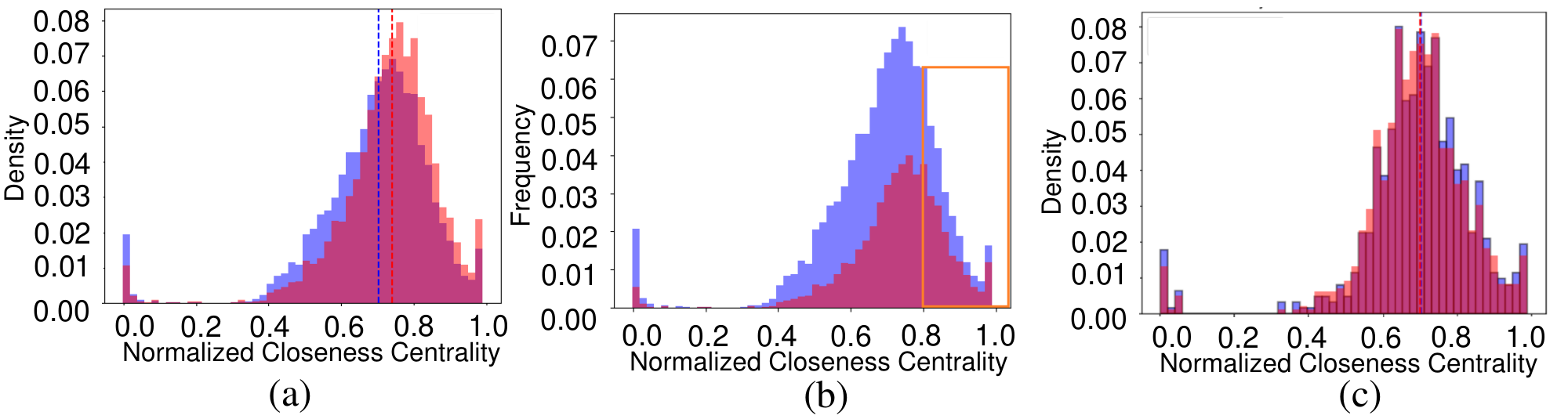}
    \caption{
    Normalized closeness centrality (CC) on \textit{Digg}.
    (a)(b) CC distributions for infected (blue) vs. source (red) nodes: (a) Density plot, with mean CC (dashed lines). (b) Frequency plot, orange box highlights CC > 0.8.
    (c) CC probability density function for ground truth (blue) vs. predicted (red) sources.}
\vspace{-1em}
    
    \label{fig:cc}
\end{figure}

The assumption of source centrality is validated by our empirical analysis of the Digg dataset. Figure~\ref{fig:cc}(a) shows the mean normalized closeness centrality (CC) of source nodes is markedly higher than other infected nodes. Furthermore, Figure~\ref{fig:cc}(b) reveals that sources constitute over 63\% of nodes with CC scores above 0.8, highlighting their key structural role.

Our proposed SIDSL model is informed by this centrality principle but uses a Graph Neural Network (GNN) for adaptive learning. As shown in Figure~\ref{fig:cc}(c), the CC probability density of SIDSL's predicted sources on Digg (mean 0.7020, std 0.1444) closely matches that of the ground truth sources (mean 0.7044, std 0.1567). This statistical alignment shows our model captures empirical source distributions without introducing bias, demonstrating a robust synthesis of knowledge-guided inference and data-driven learning.


\begin{acks}
This work was supported by the Shenzhen Ubiquitous Data Enabling Key Lab under grant ZDSYS20220527171406015, the Tsinghua Shenzhen International Graduate School-Shenzhen Pengrui Endowed Professorship Scheme of the Shenzhen Pengrui Foundation and the National Natural Science Foundation of China under U23B2030.
\end{acks}

\section*{GenAI Usage Disclosure}

The authors used Claude (Anthropic) to assist with specific formatting tasks and language editing in this work. Specifically, Claude was used to:
\begin{itemize}
    \item Convert pre-existing tabular data into LaTeX table format for presentation purposes
    \item Provide grammar correction and language polishing of author-written text
\end{itemize}

No content generation, analysis, interpretation, or creation of new ideas was performed using generative AI tools. All original research, methodology, results, and conclusions presented in this work are entirely the authors' own. The use of Claude was limited to technical formatting assistance and language refinement, similar to using grammar checking tools or word processing software.


\bibliographystyle{ACM-Reference-Format}
\balance
\bibliography{sample-base}










\end{document}